\documentstyle[11pt,twoside,sympo2011,epsfig,graphicx]{article}
\pdfoutput=1
\begin{document}
\title{Investigation of three red giants observed in the CoRoT seismo field}
\author{S. Hekker$^{1,2}$, T. Morel$^3$, A. Mazumdar$^4$, F. Baudin$^5$, E. Poretti$^6$, M.Rainer$^6$}
\affil{$^1$Astronomical Institute `Anton Pannekoek', University of Amsterdam, Science Park 904, 1098 HX Amsterdam, the Netherlands [S.Hekker@uva.nl], $^2$School of Physics and Astronomy, University of Birmingham, Edgbaston, Birmingham B15 2TT, UK, $^3$Institut d'Astrophysique et de G\'eophysique, Universit\'e de Li\`ege, All\'ee du 6 Ao\^ut, 4000 Li\`ege, Belgium, $^4$Homi Bhabha Centre for Science Education, TIFR, V. N. Purav Marg, Mankhurd, Mumbai 400088, India, $^5$Institute d'Astrophysique Spatiale, UMR 8617, Universit\'e Paris XI, B\^atiment 121, 91405 Orsay Cedex, France, $^6$INAF - Osservatorio Astronomico di Brera, via E. Bianchi 46, 23807, Merate (LC), Italy} 
\begin{abstract}
Three red giants (HD 49566 (G5III), HD 169370 (K0III) and HD 169751 (K2III)) have been observed in the CoRoT seismo field and additional ground-based spectra have been acquired. We present preliminary results of a detailed study of these stars using the observational constraints from the spectra and CoRoT data, and models from the YREC stellar evolution code.
\end{abstract}
\section{Data}
HD 49566 has been observed for about 25 days during SRa01. HD 169370 and HD 169751 have been observed for three months during LRc03. Global oscillation parameters $\nu_{\rm max}$ (frequency of maximum oscillation power) and $\Delta\nu$ (frequency separation between modes of the same order and consecutive degrees) have been computed using the methods described by Hekker et al. (2010a). Also individual frequencies have been derived by fitting the oscillation modes (Hekker et al. 2010b). 

Stellar parameters of the three stars are listed in Table~\ref{param}. For more details about the determination of these values we refer to Morel et al. 2011 (these proceedings).

\begin{table}[h] 
\begin{center}
\caption{Apparent magnitude ($m_{\rm V}$), effective temperature ($T_{\rm eff}$), surface gravity ($\log(g)$) from spectroscopy (a) and from asteroseismology (b, Miglio private communication), metallicity ([Fe/H]), parallax (plx), frequency of maximum oscillation power ($\nu_{\rm max}$) and mean large frequency separation between modes of the same degree and consecutive orders ($\Delta \nu$) of the three stars under detailed study.
}            
\begin{tabular}{lccccccc}       
\hline\hline                 
\small{star} & \small{$m_{\rm V}$}  & \small{$T_{\rm eff}$ [K]} & \small{$\log(g)$ (c.g.s.)} &  \small{[Fe/H]} & \small{plx [mas]} & \small{$\nu_{\rm max}$ [$\mu$Hz]} & \small{$\Delta\nu$ [$\mu$Hz]}\\
\hline
\small{HD 49566} & \small{7.71} & \small{5170$\pm$75} & \small{3.01$\pm$0.17$^a$} & \small{$-$0.04$\pm$0.07} & \small{~3.54$\pm$0.67} & \small{95.0$\pm$3.5} & \small{7.23$\pm$0.16}\\
	&  &	\small{5185$\pm$50} & \small{2.889$\pm$0.021$^b$} &	 \small{$-$0.04$\pm$0.06} & & & \\			
\small{HD 169370} & \small{6.31} & \small{4520$\pm$85} & \small{2.31$\pm$0.22$^a$} & \small{$-$0.27$\pm$0.11} & \small{10.52$\pm$0.57} & \small{27.2$\pm$1.1} & \small{3.36$\pm$0.11}\\
	& &	\small{4520$\pm$60} & \small{2.322$\pm$0.021$^b$} & \small{$-$0.26$\pm$0.07} & & & \\		
\small{HD 169751} & \small{8.37} & \small{4900$\pm$80} & \small{2.72$\pm$0.19$^a$} & \small{~~0.00$\pm$0.09} & \small{~7.26$\pm$0.75} & \small{61.1$\pm$2.4} & \small{5.68$\pm$0.13}\\
	& & \small{4910$\pm$55} & \small{2.683$\pm$0.026$^b$} & \small{~~0.00$\pm$0.07} & & & \\			
\hline\hline
\end{tabular}
\label{param}      
\end{center}
\end{table}

\begin{figure}
\begin{center}
\begin{minipage}{0.65\linewidth}
\centering
\includegraphics[width=\linewidth]{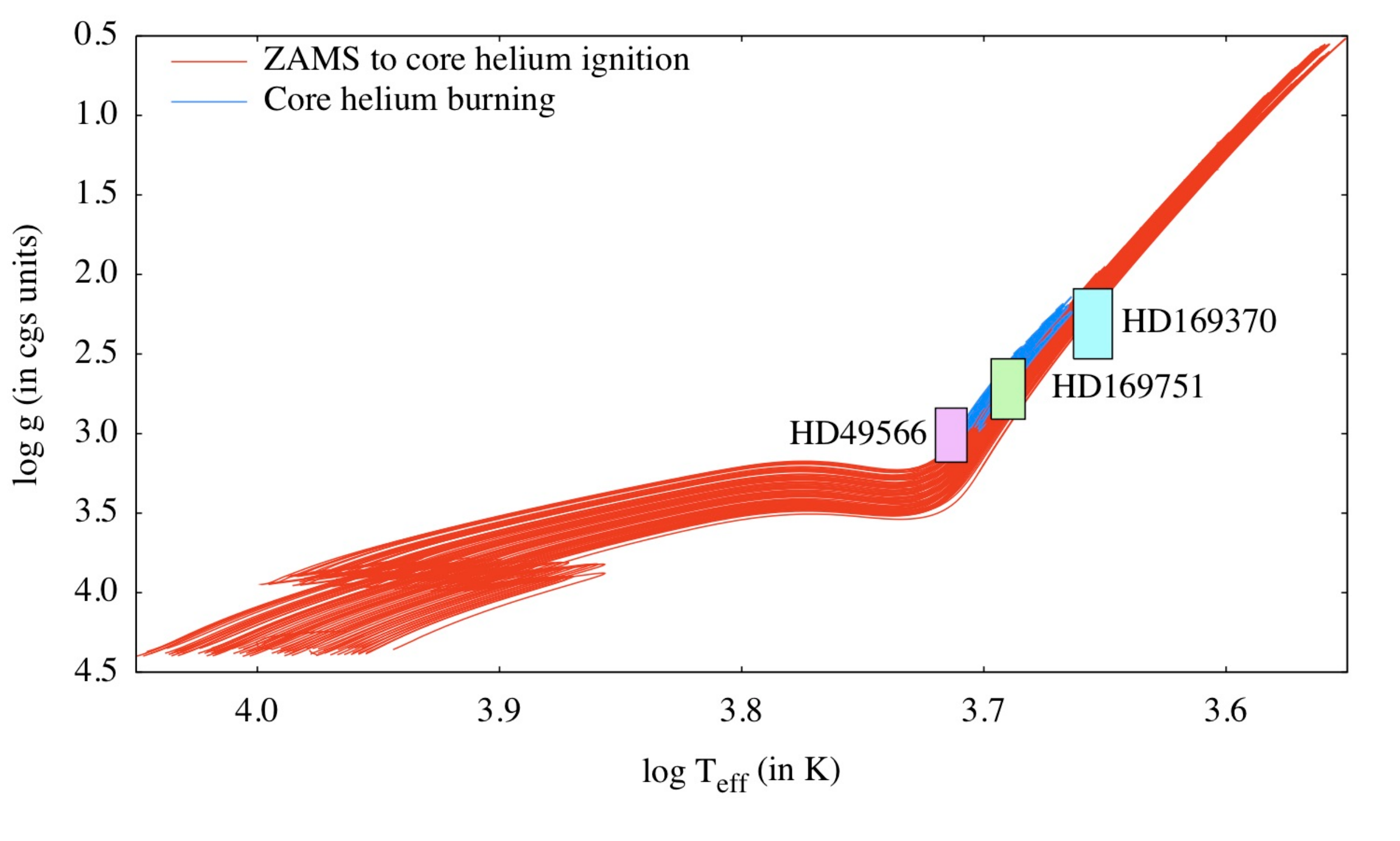}
\end{minipage}
\caption{Stellar evolutionary tracks created with YREC are shown along with spectroscopic error boxes in the $\log(g)$ - $\log T_{\rm eff}$ plane for the three CoRoT red giants. The red tracks show the evolution from the zero age main sequence till helium ignition in the core, while the blue tracks represent models in the core helium burning phase. The models span the metallicity range of $Z_0$ = [0.012,0.017] and the mass range of M/M$_{\odot}$ = [1.8,2.4]. Convective core overshoot in the main sequence phase has not been included.}
\label{models}
\end{center}
\end{figure}

\section{Modelling}
We have constructed theoretical models of red giants in both shell H-burning and core He-burning phases with the YREC code (Demarque et al. 2008). These models use the OPAL equation of state (Rogers \& Nayfonov 2002) and opacities (Iglesias \& Rogers 1996). Low temperature opacities are taken from Ferguson et al. (2005). The nuclear reaction rates from Bahcall \& Pinnsoneault (1992) are used. Convection is described by the standard mixing length theory with $\alpha$ = 1.8 H$_{\rm p}$. No convective overshoot or diffusive mixing is considered. 

The models partially cover the range of metallicities estimated for the three CoRoT red giants (see Figure~\ref{models}). The mass range is limited to M/M$_{\odot}$ = 1.8 to 2.4. Work is in progress to cover a wider range of mass, metallicity, mixing length and overshoot parameters.

We find that HD169751 can be in either the shell H-burning or core He-burning phase. The error boxes of the other two stars mostly overlap shell H-burning tracks. However, neither possibility can be ruled out before exploring the full set of model parameters. The comparison of individual frequencies and seismic parameters like $\Delta\nu$ and $\nu_{\rm max}$ should help to determine the evolutionary phase or the stellar parameters (e.g., Bedding et al. 2011).

\end{document}